\documentclass[twocolumn]{aastex62}
\usepackage{enumerate}

\newcommand{\lMsMs}{{${\rm log}\,h^{2}M_*/M_{\odot}$}}
\newcommand{\Mpc}{{$h^{-1}$Mpc}}

\begin{document}

\title{Properties of Galaxies in Cosmic Filaments around the Virgo Cluster}
\shorttitle{Lee et al.}

\correspondingauthor{Youngdae Lee; Soo-Chang Rey}
\email{hippo206@gmail.com; screy@cnu.ac.kr}

\author[0000-0002-6261-1531]{Youngdae Lee} 
\affil{Department of Astronomy and Space Science, Chungnam National University, Daejeon 34134, Republic of Korea}

\author[0000-0003-3474-9047]{Suk Kim} 
\affil{Department of Astronomy and Space Science, Chungnam National University, Daejeon 34134, Republic of Korea}

\author[0000-0002-0041-6490]{Soo-Chang Rey} 
\affil{Department of Astronomy and Space Science, Chungnam National University, Daejeon 34134, Republic of Korea}

\author[0000-0003-0469-345X]{Jiwon Chung} 
\affil{Korea Astronomy and Space Science Institute 776, Daedeokdae-ro, Yuseong-gu, Daejeon 34055, Republic of Korea}

\begin{abstract}

We present the properties of galaxies in filaments around the Virgo cluster with respect to their vertical distance from the filament spine using the NASA-Sloan Atlas catalog. The filaments are mainly composed of low-mass, blue dwarf galaxies. We observe that the $g-r$ color of galaxies becomes blue and stellar mass decreases with increasing vertical filament distance. The galaxies were divided into higher-mass ({\lMsMs}$ > 8$) and lower-mass ({\lMsMs}$\leq 8$) subsamples. We also examine the $g-r$ color, stellar mass, H$\alpha$ equivalent width (EW(H$\alpha$)), near-ultraviolet ($NUV$)$-r$ color, and HI fraction distributions of the two subsamples against the vertical distance. The lower-mass galaxies exhibit a negative $g-r$ color gradient, whereas higher-mass galaxies have a flat $g-r$ color distribution. We observe a negative EW(H$\alpha$) gradient for higher-mass galaxies, whereas lower-mass galaxies show no distinct EW(H$\alpha$) variation. In contrast, the $NUV-r$ color distribution of higher-mass galaxies shows no strong trend, whereas the lower-mass galaxies show a negative $NUV-r$ color gradient. We do not witness clear gradients of HI fraction in either the higher- or lower-mass subsamples. We propose that the negative color and stellar mass gradients of galaxies can be explained by mass assembly from past galaxy mergers at different vertical filament distances. In addition, galaxy interactions might be responsible for the contrasting features of EW(H$\alpha$) and $NUV-r$ color distributions between the higher- and lower-mass subsamples. The HI fraction distributions of the two subsamples suggest that ram-pressure stripping and gas accretion could be ignorable processes in the Virgo filaments.

\end{abstract}


\section{Introduction}

In the frame of the hierarchical structure formation in the standard cold dark matter universe, the build-up of galaxy clusters is characterized by the accretion of galaxies into higher-density cluster environments through filamentary structures \citep{Bon96,Spr05}. It has been proposed that the filament environment around galaxy clusters is closely linked to galaxy evolution in terms of the subject of ``pre-processing" \citep{Fuj04}, where properties of galaxies may already have been altered before they enter galaxy clusters \citep{Zab96,DeL12,Mah13,Cyb14}. Therefore, filaments are ideal places for the investigation of the physical processes that control the transitioning of galaxies in less-dense environments to cluster galaxies.

Recently, the specific role of pre-processing in filaments has been extensively explored in different surveys. In particular, there is a growing body of observational evidence indicating that the properties of galaxies (e.g., color, mass, red galaxy fraction, and star-formation rate) change as a function of their distance from the filament spine \citep[e.g.,][]{Mar16,Alp16,Che17,Kuu17,Mal17,Mah18,Kra18,Bon19,Lub19,Sar19}. Galaxies close to filaments show a tendency to have redder colors, lower H$\alpha$ emission-line equivalent width (EW(H$\alpha$)), and higher early- to late-type galaxy fraction, indicating that the efficiency of star-formation quenching varies with distance from filaments. Furthermore, higher-mass galaxies are preferentially located closer to filaments. These observational results support the idea that galaxies could be effectively pre-processed in a moderately dense filament environment.

Several environmental effects are proposed as possible physical mechanisms that could be responsible for the galaxy pre-processing in filaments. Merger and tidal interaction between galaxies are expected to be a prevailing process in the filament environment \citep{Dar15,Mal17,Kra18,Mah18}. In addition, stripping of gas within galaxies in filaments can take place by warm-hot intergalactic medium (WHIM; $10^5$ K $<$ T $<$ $10^7$ K, e.g., \citealt{Ben13}; see also \citealt{Ton07} for ram-pressure stripping in the cluster periphery) within filaments. As another process, galaxies moving along the filaments can accrete cold gas from intrafilament medium \citep{Ker05,San08,Dar15,Kle17}.

Up to date, most observational studies of filaments have concentrated on massive galaxies with $>$10$^{10} M_{\odot}$ mainly due to that filaments are only defined at intermediate and high redshifts. On the other hand, since low-mass ($<$10$^9 M_{\odot}$) dwarf galaxies, with low binding energies, are more susceptible to even weak perturbations than massive galaxies, they are a better tool to probe the details of the multiple processes occurring in filaments that can affect the properties of filament galaxies. Recently, \citet{Kim16} identified seven filaments around the Virgo cluster, the nearest rich and dynamically young cluster \citep{Agu05}, in which the majority of galaxies are faint ($M_B > -19$ mag; $\sim$88\% of the total sample) dwarfs. Therefore, Virgo filaments are prime targets for detailed investigations of the physical processes affecting galaxy properties in filament environment.

In response to this situation, in this work, we explored the impact of the filaments in the vicinity of the Virgo cluster on properties of filament galaxies using the photometric and spectroscopic data of the Sloan Digital Sky Survey (SDSS), Galaxy Evolution Explorer (GALEX) ultraviolet (UV) data, and neutral hydrogen (HI) gas data of the Arecibo Legacy Fast ALFA (ALFALFA) blind survey. In Section~\ref{sec:data}, we describe the basic data of galaxies in and around the Virgo filaments. Section~\ref{sec:mem} describes the selection of member galaxies in filaments. In Section~\ref{sec:results}, we show our results on the variations of galaxy properties (optical and UV colors, mass, central and global star formations, and HI gas content) as a function of the vertical filament distance. A discussion on physical processes in filaments is given in Section~\ref{sec:discussion}. Finally, we summarize our results in Section~\ref{sec:summary}. Throughout this paper, we assume a $h = H_0/(100\,{\rm km}\,{\rm s}^{-1}\,{\rm Mpc}^{-1})$, the matter density $\Omega_m = 0.27$, and the dark energy density $\Omega_{\lambda} = 0.73$ \citep{Kom11}.

\section{Data} \label{sec:data}

Our basic sample is drawn from galaxies at $z < 0.014$ in and around the Virgo filaments from the NASA-Sloan Atlas (NSA) catalog. In order to identify galaxies associated with galaxy groups residing on filaments, we make use of our sophisticated galaxy group catalog (Lee et al. in prep.). This is constructed for SDSS galaxies in the redshift range of $0.001 < z < 0.2$ with the use of the friends-of-friends for group detection. We examined the radial velocity distribution of NSA galaxies as a function of the group-centric distance. In this distribution, we overplotted a spherical symmetric infall model using measured velocity dispersion and virial radius of each group \citep{Pra94}. We assigned NSA galaxies to group galaxies if they are bounded by the caustic curves of the infall model. Finally, we only selected galaxies that do not belong to groups in order to investigate how galaxies change their properties with varying distances from the filaments, avoiding contamination from member galaxies of groups located in filaments \citep[e.g.,][]{Alp16,Bon19}. This allows studying the galaxy evolution in filament unaffected by local environmental effects occurring in groups \citep{Rob13,Dav15,Alp16}.

\begin{figure*}
\epsscale{1.2}
\plotone{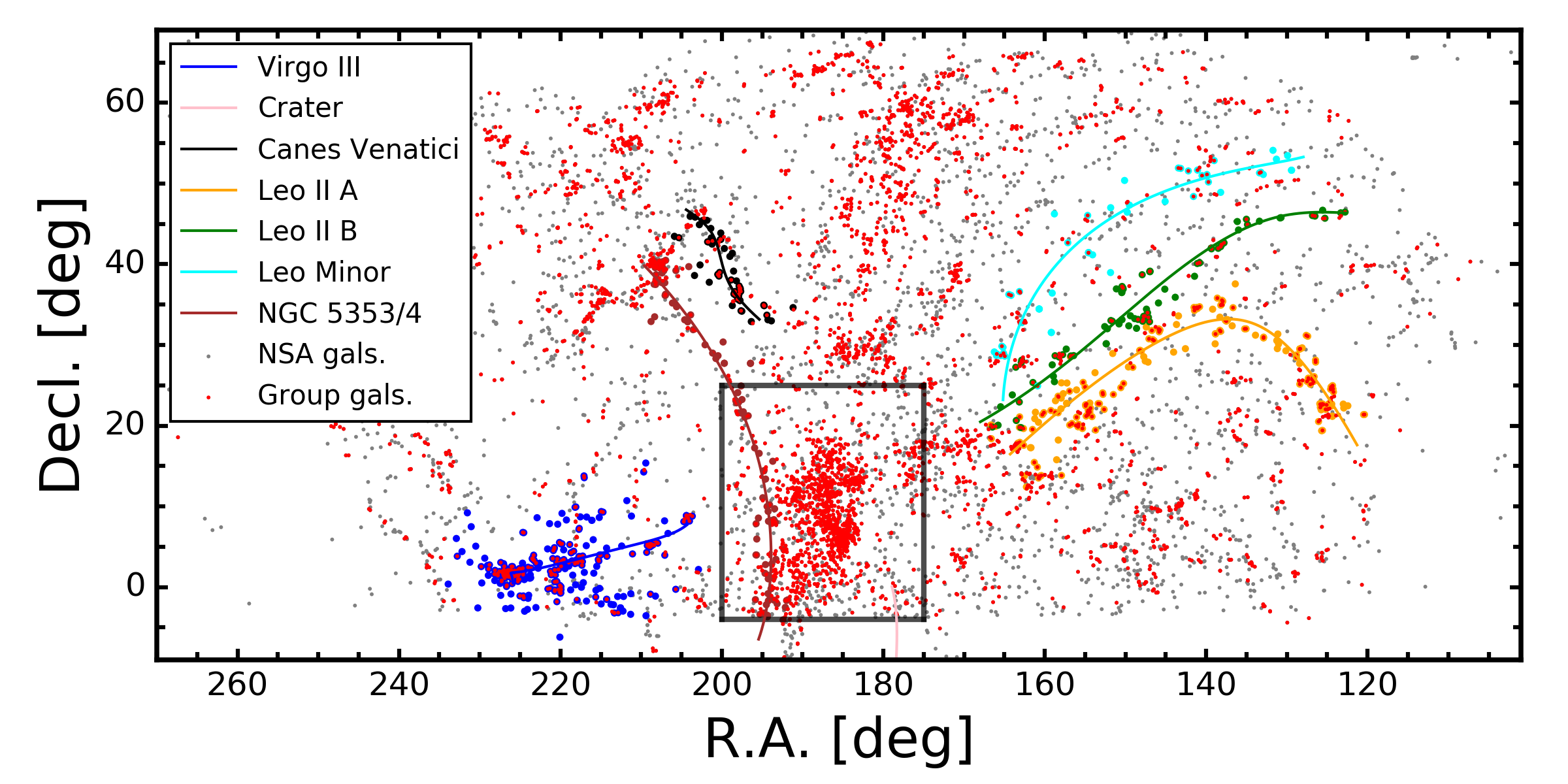}
\caption{Spatial distribution of galaxies around the Virgo filaments in the equatorial coordinate system. Colored curves are spines of seven filaments identified by \citet{Kim16} and colored circles are selected member galaxies that belong to the filaments. NSA galaxies at $z < 0.014$ and group galaxies are denoted as gray dots and red circles, respectively. The large rectangular box is the region of the Virgo cluster covered by the Extended Virgo Cluster Catalog \citep[EVCC,][]{Kim14}. In our study, we only consider galaxies that do not belong to galaxy groups (colored circles without red dots).
}
\label{NSASpatial}
\end{figure*}

Figure~\ref{NSASpatial} shows the sky distribution of galaxies around the filaments. Gray dots indicate NSA galaxies at $z < 0.014$; different colored circles are the member galaxies belonging to different filaments (see Sec.~\ref{sec:mem}), and red circles are the group galaxies. The spines of different filaments are plotted with different colored curves. The Crater filament galaxies are excluded from our analysis since the NSA catalog does not cover most of this filament in the southern hemisphere.

\section{Selection of Member Galaxies in Filaments} \label{sec:mem}

Since our study focuses on the properties of filament galaxies in the vertical direction from the filament spine, we need to define possible member galaxies belonging to each filament structure. In order to construct an accurate spatial distribution of galaxies in the NSA, we first converted the observed heliocentric radial velocities extracted from the NSA into velocities relative to the centroid of the Local Group using a method from \citet{Kar96}. All NSA galaxies were then mapped in the supergalactic coordinate system (SGX, SGY, SGZ), which was converted from heliocentric radial velocities and positions (R.A. and Decl.) of galaxies based on the assumption of a linear relationship between radial velocity and distance. We adopted the spines of filaments obtained by \citet{Kim16} which are created by the three-dimensional third-order polynomial fitting of visuially selected filament structures (see \citealt{Kim16} for details). We finally measured the three-dimensional vertical distances of galaxies ($D_{\rm ver}$) from the filament spine in Mpc unit. In the case of Virgo III filament, since it is contiguous to the Virgo cluster in the SGZ direction, we rejected galaxies within a distance of 1.5 {\Mpc} from the center of the supergalactic coordinate system in SGZ direction to avoid contamination from cluster galaxies.

\begin{figure}
\epsscale{1.2}
\plotone{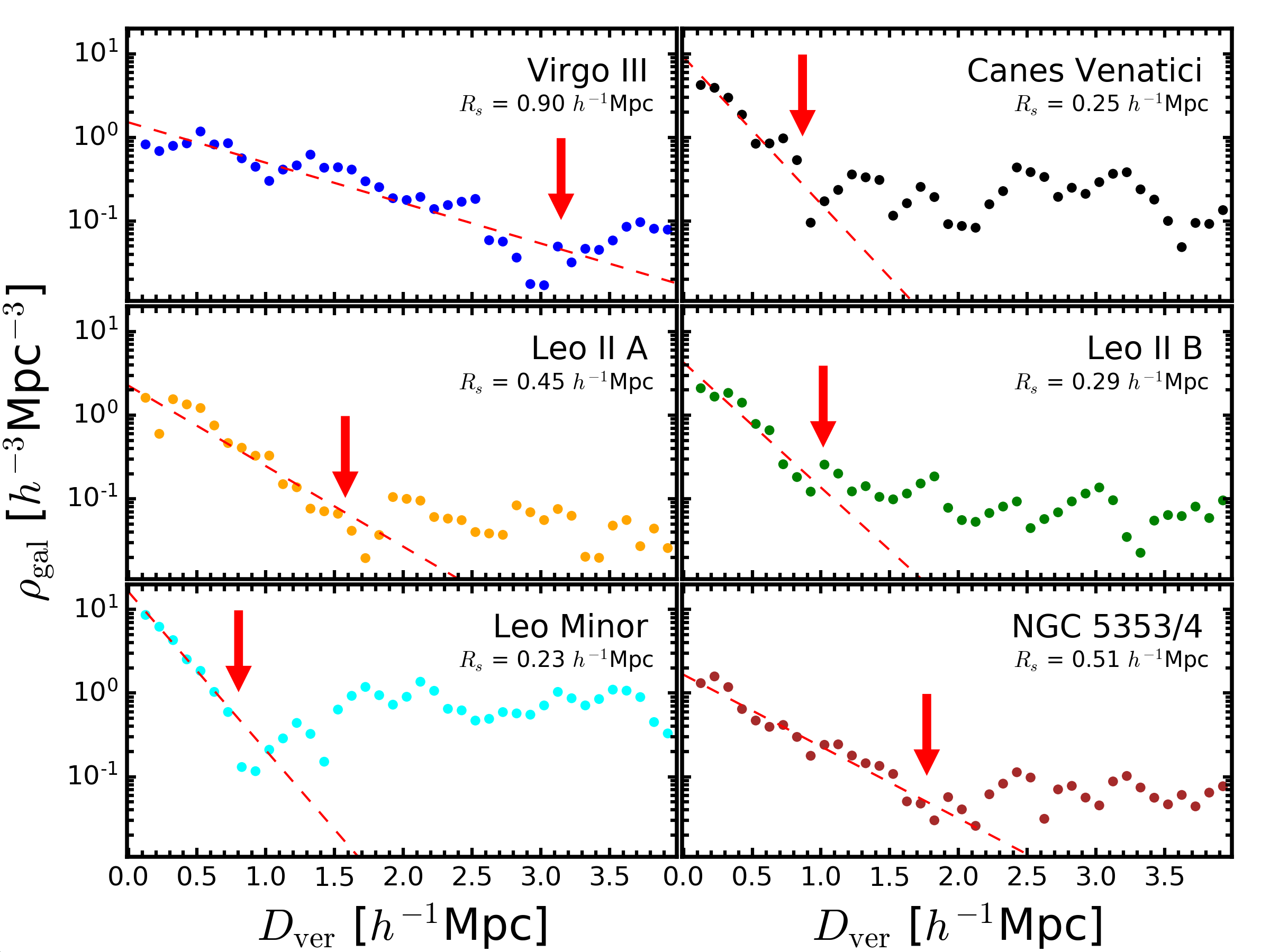}
\caption{Galaxy number density profiles of six filaments as a function of the vertical distance from the filament spine ($D_{\rm ver}$). Circles are observed number densities at different $D_{ver}$ and dashed lines denote the best-fit exponential models. In each panel, scale length ($R_s$) of each filament is indicated and filament membership criterion (i.e., within 3.5$R_s$ from the filament spine) is presented as a red arrow.
}
\label{VerDen}
\end{figure}

Figure~\ref{VerDen} presents galaxy number density ($\rho_{\rm gal}$) profiles of each filament as a function of $D_{\rm ver}$. We define $\rho_{\rm gal}$ at a $D_{\rm ver}$ as the number of galaxies over cylindrical volume along the filament spine in the supergalactic coordinate system. The $\rho_{\rm gal}$ was measured by the moving bin method with a bin size of 0.25 {\Mpc} and a step size of 0.1 {\Mpc} in $D_{\rm ver}$ direction. The overall feature in all filaments is that $\rho_{\rm gal}$ systemically decreases from the filament spine to a certain $D_{\rm ver}$ and then shows no variation with increasing $D_{\rm ver}$. We have fitted decreasing part of the observed $\rho_{\rm gal}$ distribution with an exponential function:
\begin{equation}
\rho_{\rm gal} = \rho_{0}exp(-D_{\rm ver}/R_s)
\end{equation}
, where $\rho_{0}$ and $R_s$ are central galaxy number density and scale length, respectively. The best-fit parameters are shown in Table~\ref{tab01} and the best-fit models are presented as red dashed lines in Figure~\ref{VerDen}. All $R_s$ are found to be less than 1 {\Mpc}. Finally, we define member galaxies belonging to the filament as those within 3.5$R_s$ from the filament spine. We note that the 3.5$R_s$ well represents the location of a local minimum density in $\rho_{\rm gal}$ distribution (see red arrows for indicating 3.5$R_s$ in Figure~\ref{VerDen}). The selected member galaxies in six filaments include a total of 289 galaxies.

For our analysis, we extracted photometric and spectroscopic parameters of member galaxies from the NSA; optical and UV magnitudes, stellar mass, and intensities of various emission lines (e.g., H$\alpha$, H$\beta$, [OIII], [NII]). These parameters in the NSA are mainly derived from the SDSS optical data and with the addition of the GALEX UV data.

\begin{table*} 
\renewcommand{\thetable}{\arabic{table}}
\centering
\caption{Properties of filaments} \label{tab01}
\begin{tabular}{l c c c c}
\hline
\hline
 Name           &       $\rho_0$              &     $R_s$             & Length         &  $N$     \\ 
                &    [$h^{-3}$Mpc$^{-3}]$   &   [$h^{-1}$Mpc]         &  [$h^{-1}$Mpc] &       \\  
\hline
Virgo III      &   1.51 $\pm$ 0.21              &  0.90 $\pm$ 0.07    &  12.39         &   92    \\
Canes Venatici &   9.09 $\pm$ 2.45             &  0.25 $\pm$ 0.03    &  7.23          &   26    \\            
Leo II A       &   2.26 $\pm$ 0.42              &  0.45 $\pm$ 0.05    &  18.91         &   57    \\          
Leo II B       &   4.27 $\pm$ 1.04              &  0.29 $\pm$ 0.04    &  16.96         &   34    \\          
Leo Minor      &   16.31 $\pm$ 4.46             &  0.23 $\pm$ 0.03    &  5.91          &   26    \\          
NGC 5353/4     &   1.66 $\pm$ 0.31              &  0.51 $\pm$ 0.05    &  23.22         &   54    \\          
\hline

\multicolumn{5}{l}{Notes. (1) Name of the filament} \\
\multicolumn{5}{l}{      (2) Central galaxy number density of the exponential best-fit model} \\
\multicolumn{5}{l}{      (3) Scale length of the exponential best-fit model} \\
\multicolumn{5}{l}{      (4) Length of the filament in the SGX-SGY-SGZ plane} \\
\multicolumn{5}{l}{      (5) Number of member galaxies that do not belong to groups} \\
\end{tabular}
\end{table*}


\section{Results} \label{sec:results}
\subsection{Stellar Mass Distribution and Color-Magnitude Diagram}

\begin{figure}
\epsscale{1.2}
\plotone{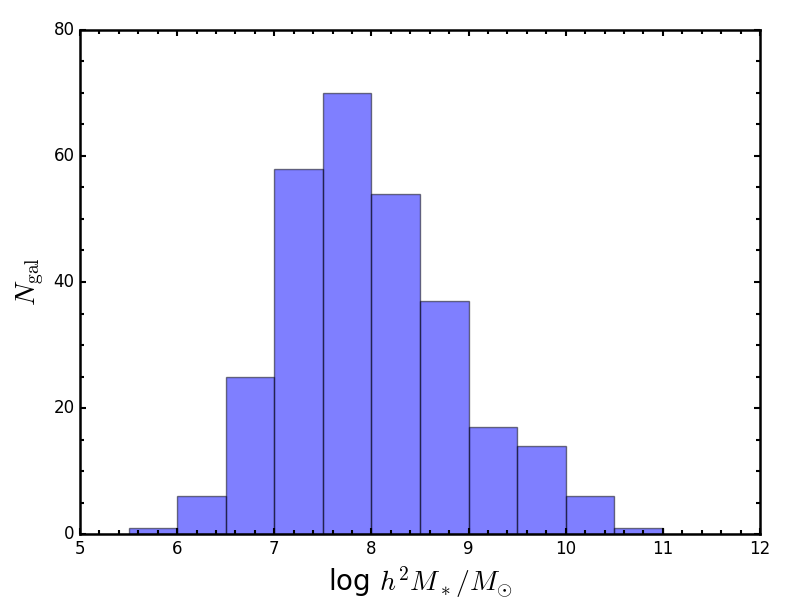}
\caption{Stellar mass distribution of member galaxies in filaments.
}
\label{MassDist}
\end{figure}

The filaments around the Virgo cluster defined by \citet{Kim16} are the nearest large-scale structures from us ($\sim$14--41 Mpc), mostly consisting of faint dwarf galaxies ($M_B > -19$; $\sim$88\% of the total sample). In Figure~\ref{MassDist}, we present the stellar mass ($M_*$) distribution of member galaxies in filaments. It is clear that most (87\%) of our filament galaxies are lower mass ones with {\lMsMs}$\,< 9$  in which median and standard deviation are 7.7 and 0.6, respectively.

We examined the photometric completeness of our galaxy sample by fitting the conventional Schechter function \citep[see Equation 2 of][]{Vul13} to the stellar mass distribution of higher-mass galaxies with {\lMsMs}$\,> 7.5$. We estimated the low-mass end slope ($\alpha = -1.34$) and the characteristic stellar mass ({\lMsMs} = 10.47) of the fitted mass function. Assuming this mass function is photometrically complete, we calculated the completeness of our filament galaxy sample at each mass bin which is the number ratio between the fitted Schechter function and observed stellar mass distribution. The completeness of galaxy samples for {\lMsMs}$\,>7$ and {\lMsMs}$\,<7$ are found to be 84\% and 4\%, respectively. We found that no significant change in our main results when we use only a relatively more complete sample with {\lMsMs}$>7$. We henceforth utilized the total sample galaxies without taking completeness into the consideration for our analysis.

\begin{figure}
\epsscale{1.2}
\plotone{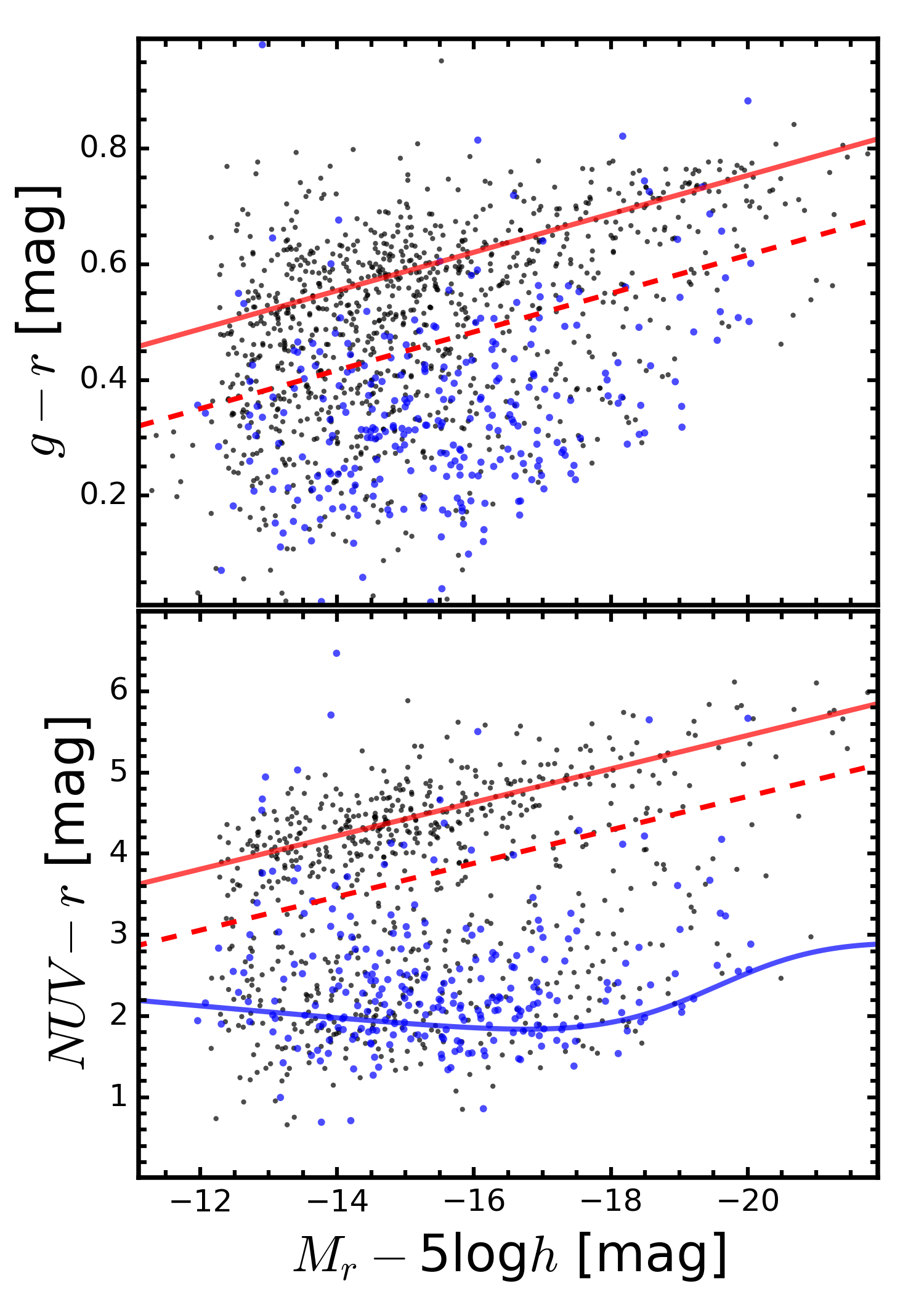}
\caption{$g-r$ (top) and $NUV-r$ (bottom) color-magnitude diagrams of galaxies in filaments (blue circles). For comparison, galaxies in the Virgo cluster are also overplotted (black dots). The red solid lines represent the red sequence of the Virgo cluster and the red dashed lines are $-3\sigma$ deviation from the red sequence. In the $NUV-r$ CMD, the blue curve indicates the blue cloud defined by \citet{Wyd07}.
}
\label{UVCMR}
\end{figure}

Figure~\ref{UVCMR} shows the $g-r$ and near-ultraviolet $(NUV)-r$ color-magnitude diagrams (CMDs) of galaxies in filaments (blue circles). For comparison with galaxies in a denser environment, we also overplot galaxies in the Virgo cluster (black dots) based on the SDSS optical and GALEX UV photometry of the EVCC catalog \citep{Kim14} adopting a distance modulus of the Virgo cluster of 31.1 \citep{Mei07}. Further, we construct red sequences in the $g-r$ and $NUV-r$ CMDs by a linear least-square fitting to the early-type galaxies in the Virgo cluster (red solid lines in Figure~\ref{UVCMR}). The filament galaxies show a blueward offset from the red sequence of the Virgo cluster in $g-r$; 80\% (232/289) lie blueward of the $-3\sigma$ deviation (red dashed line) from the Virgo red sequence in $g-r$. A more distinct offset is shown in the $NUV-r$ CMD owing to the longer baseline of $NUV-r$ color; 81\% (234/289) of the filament galaxies lie blueward of the $-3\sigma$ deviation (red dashed line) from the red sequence of the Virgo cluster. The $NUV-r$ CMD evidently shows that, at all magnitudes, the majority of filament galaxies occupy the region of the blue cloud as defined by galaxies in the local Universe \citep[][blue curve]{Wyd07}. The $NUV-r$ CMD is particularly efficient to trace recent star-formation activity since the UV flux is sensitive to young ($<$1 Gyr) stellar populations. This indicates that the Virgo filaments are dominated by galaxies that have experienced recent star-formation.

\subsection{Trends in Color and Stellar Mass}

Gradients of color and stellar mass of galaxies are found in observed filaments consisting of massive galaxies with $>$$10^{10}$ $M_{\odot}$ ; redder and more massive galaxies are closer to filaments, whereas the bluer and less massive galaxies are found further away from the filaments \citep[e.g.,][]{Alp16,Mal17,Kra18,Bon19,Lub19,Sar19}. It is intriguing to investigate whether low-mass dwarf galaxies in filaments also follow this trend.

We examined color and stellar mass distributions of filament galaxies as a function of the scaled vertical distance from the filament spine ($D_{\rm ver}$/$R_s$) which is defined as $D_{\rm ver}$ divided by the measured $R_s$ of the filament. Further, we calculated mean color and mean mass along the $D_{\rm ver}$/$R_s$ using the moving bin method, where bin size and step size of the $D_{\rm ver}$/$R_s$ are 1 and 0.4, respectively. The errors of color and mass of each bin are calculated by the bootstrap resampling method.

\begin{figure}
\epsscale{1.2}
\plotone{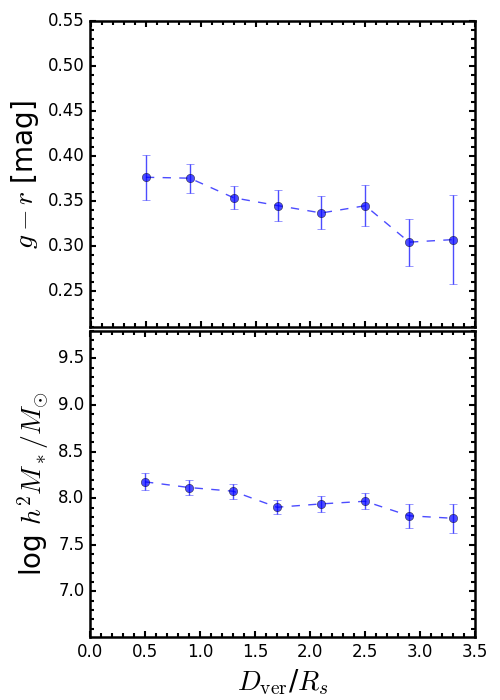}
\caption{$g-r$ color (top) and stellar mass (bottom) versus scaled vertical distance from the filament spine ($D_{\rm ver}$/$R_s$). The error bar of each bin indicates bootstrap resampling uncertainty.
}
\label{ColorVDist}
\end{figure}

Figure~\ref{ColorVDist} illustrates $g - r$ color (top) and stellar mass (bottom) distributions as a function of the $D_{\rm ver}$/$R_s$. It appears that color and stellar mass correlate with the vertical distance from the filament spine; that is, $g-r$ color becomes blue and stellar mass decreases with increasing $D_{\rm ver}$/$R_s$. We performed a Kolmogorov-Smirnov (K-S) test to quantify the statistical significance of this trend. We divided galaxies into two subsamples located in the inner ($0 \leq D_{\rm ver}/R_s \leq 1.75$) and outer ($1.75 < D_{\rm ver}/R_s \leq 3.5$) regions of the filament. The test between two subsamples provides probabilities of 0.0312 and 0.0126 for color and stellar mass, respectively, rejecting that the color and stellar mass distributions between two subsamples are drawn from the same parent one. This implies that negative gradients of color and stellar mass are evident in filaments mostly consisting of low-mass ($<$10$^9$ $M_{\odot}$) dwarf galaxies. Combining with previous results mainly based on more massive galaxies, it can be said that the color and stellar mass gradients are generally observed in the vertical direction of filaments regardless of mass range.

\begin{figure}
\epsscale{1.2}
\plotone{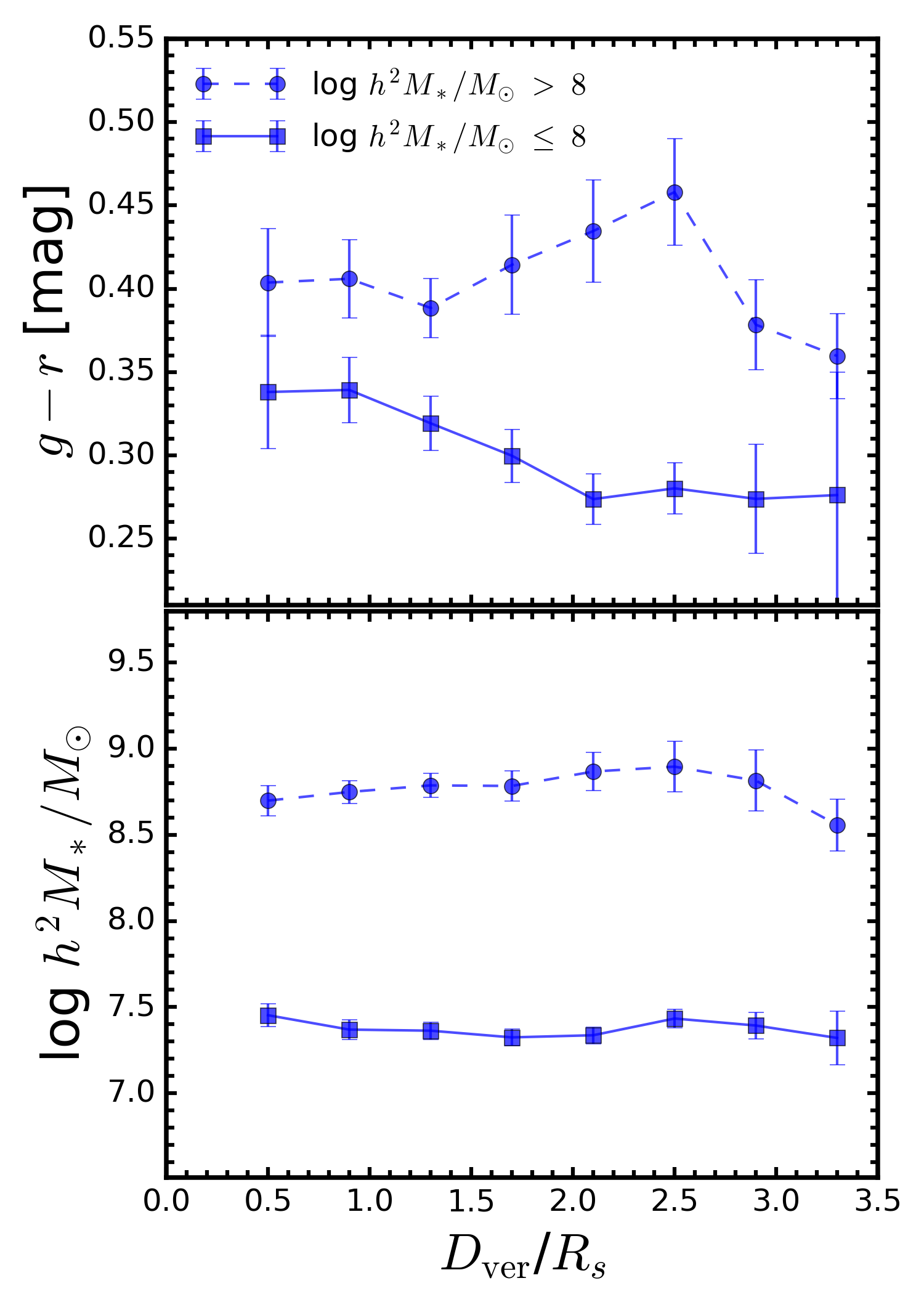}
\caption{$g-r$ color (top) and stellar mass (bottom) versus scaled vertical distance from the filament spine ($D_{\rm ver}$/$R_s$) for galaxies in different mass ranges. The circles and dashed lines represent higher-mass ({\lMsMs} $> 8$)  galaxies and the squares and solid lines are lower-mass ({\lMsMs} $\leq 8$) galaxies. The error bar of each bin indicates bootstrap resampling uncertainty.
}
\label{MassVDist}
\end{figure}

We divided filament galaxies into higher-mass ({\lMsMs} $> 8$) and lower-mass ({\lMsMs} $\leq 8$) subsamples where {\lMsMs} $= 8$ is a median stellar mass of filament galaxies. Figure~\ref{MassVDist} shows $g-r$ color and stellar mass distributions of higher-mass (circles and dashed line) and lower-mass (squares and solid line) galaxies as a function of the $D_{\rm ver}$/$R_s$. The most notable feature is that the color distribution of higher-mass galaxies appears to be different from that of lower-mass galaxies (see top panel of Figure~\ref{MassVDist}); higher-mass galaxies show a rather flat distribution, whereas the color of lower-mass galaxies appears to decrease with the $D_{\rm ver}$/$R_s$. The K-S tests of higher- and lower-mass galaxies yield probabilities of 0.8137 and 0.0450, respectively, for color distributions between the inner ($0 \leq D_{\rm ver}/R_s \leq 1.75$) and outer ($1.75 < D_{\rm ver}/R_s \leq 3.5$) regions of the filament. This indicates no statistically significant difference in the color distributions of high mass-galaxies between the inner and outer regions of the filament. On the other hand, the color distribution of lower-mass galaxies closer to the filament spine shows a statistically significant difference from that of counterparts far from the filament.

In the case of stellar mass distributions, both of the two subsamples in different mass ranges do not show a significant variation of stellar mass with the $D_{\rm ver}$/$R_s$ (see bottom panel of Figure~\ref{MassVDist}). The K-S tests of higher- and lower-mass galaxies yield probabilities of 0.3187 and 0.3314, respectively, for stellar mass distributions between the inner ($0 \leq D_{\rm ver}/R_s \leq 1.75$) and outer ($1.75 < D_{\rm ver}/R_s \leq 3.5$) regions of the filament. This indicates that the stellar mass gradients of two subsamples are not statistically significant.

\subsection{Star-forming Galaxies in Filaments}

Star-forming galaxies are mostly found in less dense environments such as field and filament \citep{Hai07,Gav10}. They are attractive objects for inspecting the degree of environmental effects in terms of their strengths of star-formation. In this regard, investigation of the star-formation activity of star-forming galaxies at different vertical distances from the filament spine will enable us to understand the environment of filament.

\begin{figure}
\epsscale{1.2}
\plotone{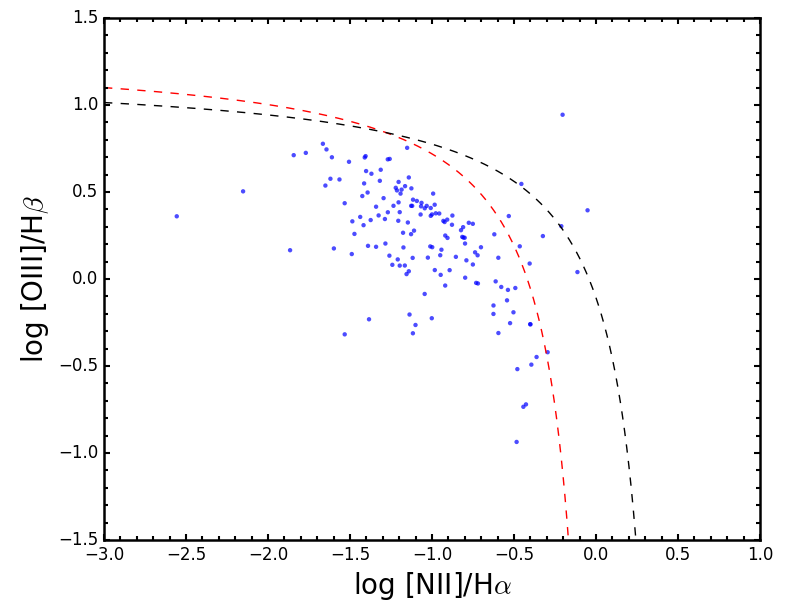}
\caption{BPT diagram of filament galaxies. Star-forming galaxies are defined as those below the red curve of \citet{Kau03}. AGNs are those above the black curve of \citet{Kew01}.}
\label{BPT}
\end{figure}

We selected star-forming galaxies using the Baldwin-Phillips-Terlevich (BPT) emission line diagnostic diagram \citep{Bal81} that allows discriminating star-forming galaxies from possible active galactic nuclei (AGNs). Figure~\ref{BPT} is the BPT diagram where the emission-line ratios [OIII]/H$\beta$ and [NII]/H$\alpha$ extracted from the NSA are used. We define star-forming galaxies which locate below the empirical demarcation line of \citet[][red curve]{Kau03}. Only galaxies with strong H$\alpha$ equivalent width (EW) (i.e., EW(H$\alpha$) $>$ 2 \AA) are also considered, and finally, 161 star-forming galaxies are identified. AGNs are located above the theoretical starburst model line of \citet[][black curve]{Kew01}.

\begin{figure}
\epsscale{1.2}
\plotone{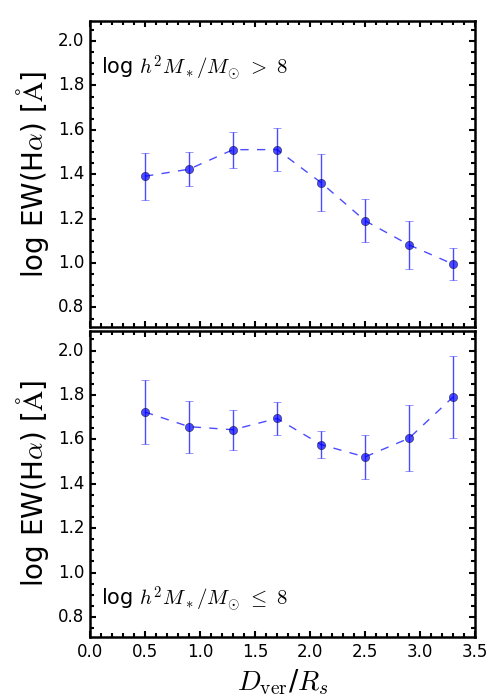}
\caption{Equivalent width of H$\alpha$ emission line (EW(H$\alpha$)) versus scaled vertical distance ($D_{\rm ver}$/$R_s$) from the filament spine of higher-mass ({\lMsMs} $> 8$) galaxies (top) and lower-mass ({\lMsMs} $\leq 8$) galaxies (bottom). The error bar of each bin indicates bootstrap resampling uncertainty.}
\label{SFGHAEW}
\end{figure}

The H$\alpha$ emission traces the ongoing star-formation activity from massive stars with timescales of a few tens of Myr. Figure~\ref{SFGHAEW} shows EW(H$\alpha$) distributions of higher-mass ({\lMsMs} $> 8$, top panel) and lower-mass ({\lMsMs} $\leq 8$, bottom panel) galaxies as a function of the $D_{\rm ver}$/$R_s$. The EW(H$\alpha$) distributions between two subsamples appear to be different; the EW(H$\alpha$) values of higher-mass galaxies likely decrease with the $D_{\rm ver}$/$R_s$, but lower-mass galaxies show no distinct trend of EW(H$\alpha$). The K-S tests of higher- and lower-mass galaxies yield probabilities of 0.0094 and 0.5685, respectively, for EW(H$\alpha$) distributions between the inner ($0 \leq D_{\rm ver}/R_s \leq 1.75$) and outer ($1.75 < D_{\rm ver}/R_s \leq 3.5$) regions of the filament. This indicates that a possible EW(H$\alpha$) gradient from the filament spine is only seen for the higher-mass galaxies. We note that EW(H$\alpha$) value from the NSA generally provides information on the central region of a galaxy due to the 3 arcsec (i.e., $204$--$596$ pc at a distance range of $14$--$41$ Mpc for our filaments) fiber aperture of the SDSS spectroscopic observations. Therefore, current star-formation in the central region of higher-mass galaxies is more active for those located closer to the filament spine.

\begin{figure}
\epsscale{1.2}
\plotone{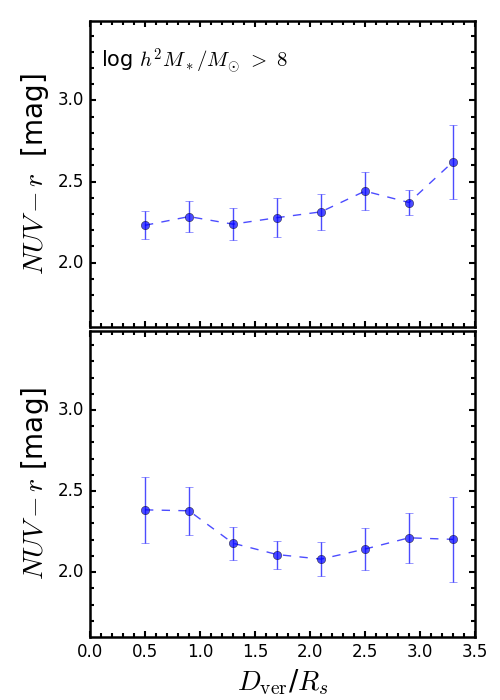}
\caption{$NUV-r$ color versus scaled vertical distance ($D_{\rm ver}$/$R_s$) from the filament spine of higher-mass ({\lMsMs} $> 8$) galaxies (top) and lower-mass ({\lMsMs} $\leq 8$) galaxies (bottom). The error bar of each bin indicates bootstrap resampling uncertainty.}
\label{SFGCD}
\end{figure}

On the other hand, the global star-formation of the entire region of a galaxy can be examined by UV photometry. Figure~\ref{SFGCD} shows $NUV-r$ color distributions of higher-mass ({\lMsMs} $> 8$, top panel) and lower-mass ({\lMsMs} $\leq 8$, bottom panel) galaxies as a function of the $D_{\rm ver}$/$R_s$. The $NUV-r$ color distribution of higher-mass galaxies is likely to show a trend with the $D_{\rm ver}$/$R_s$; $NUV-r$ color very smoothly increases with the $D_{\rm ver}$/$R_s$. On the other hand, the $NUV-r$ colors of lower-mass galaxies appear to decrease with the $D_{\rm ver}$/$R_s$. However, the K-S tests of higher- and lower-mass galaxies yield probabilities of 0.0904 and 0.0387, respectively, for $NUV-r$ color distributions between the inner ($0 \leq D_{\rm ver}/R_s \leq 1.75$) and outer ($1.75 < D_{\rm ver}/R_s \leq 3.5$) regions of the filament, indicating that $NUV-r$ color gradient is statistically significant only for lower-mass galaxies. This implies that recent ($<$1 Gyr), global star-formation of lower-mass galaxies residing closer to the filament is less active than counterparts with large displacement from the filament spine.

\subsection{HI Content of Filament Galaxies}

\begin{figure*}
\epsscale{1.2}
\plotone{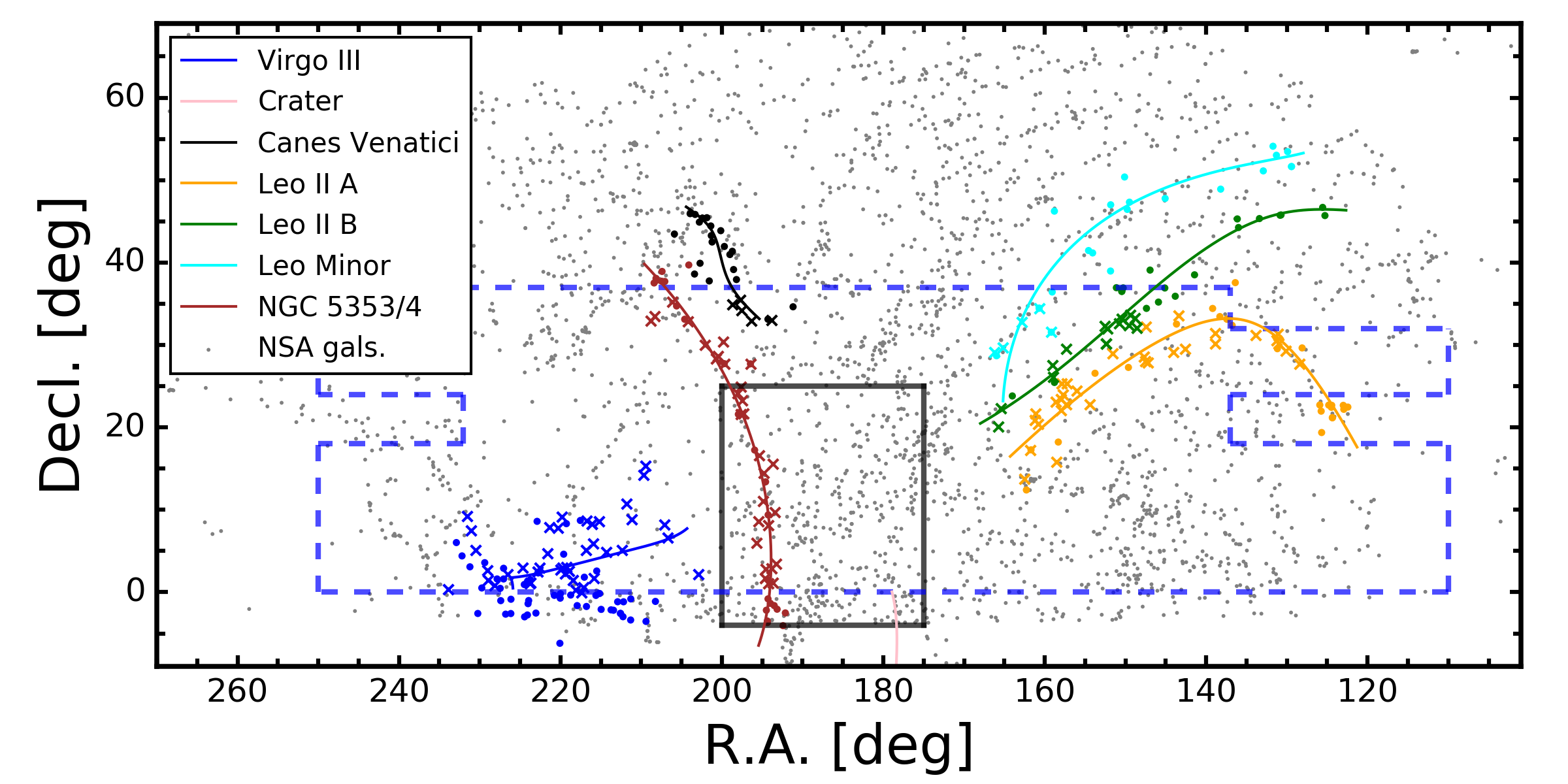}
\caption{Spatial distribution of filament galaxies with HI detection from the ALFALFA survey data (colored crosses). Colored circles denote filament galaxies with HI non-detection. Gray circles are NSA galaxies not associated with the filaments and the galaxy groups. The large rectangular box is the region of the EVCC. The coverage of the ALFALFA survey is presented by the blue dashed contour.}
\label{HISpatial}
\end{figure*}

HI gas content within a galaxy has been a good tracer of different environmental effects in a cluster environment \citep[e.g.,][]{Bos06}. In terms of the role of the filament environment in regulating the evolution of galaxies, we also investigate the HI gas content of galaxies residing on our filaments using data from the ALFALFA \citep{Hay18} blind survey. The flux limit of the ALFALFA corresponds to the detection limit of the HI mass $M_{HI} = 10^{7.4}$ $M_{\odot}$ at the distance of the Virgo cluster \citep{Hay11,Hay18}. We note that owing to proximity to us, Virgo filaments are optimal for the detection of HI gas in faint, low mass galaxies. Of our 289 filament galaxies, 184 are included in the sky coverage of the ALFALFA survey \citep[][see dashed contour in Figure~\ref{HISpatial}]{Hay18}, which covers the area around the Virgo cluster at declinations $0^{\circ}$--$37^{\circ}$. We found that HI gas of 69\% (i.e., 127 galaxies) of the galaxies are detected and HI gas masses of galaxies are also extracted from the ALFALFA data. Figure~\ref{HISpatial} shows the sky distribution of HI detected galaxies (crosses).

\begin{figure}
\epsscale{1.2}
\plotone{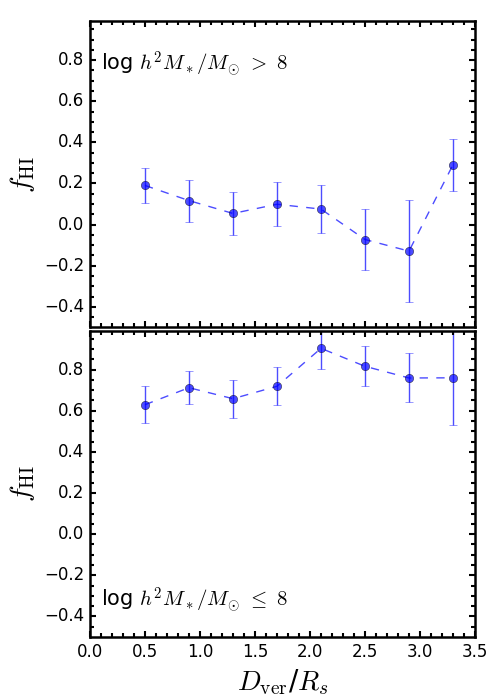}
\caption{HI gas fraction ($f_{HI}$) versus scaled vertical distance ($D_{\rm ver}$/$R_s$) from the filament spine of higher-mass ({\lMsMs} $> 8$) galaxies (top) and lower-mass ({\lMsMs} $\leq 8$) galaxies (bottom). The error bar of each bin indicates bootstrap resampling uncertainty.}
\label{HIFVDist}
\end{figure}

Figure~\ref{HIFVDist} shows the HI fraction ($f_{\rm HI}$) distributions of higher-mass ({\lMsMs} $> 8$, top panel) and lower-mass ({\lMsMs} $\leq 8$, bottom panel) galaxies as a function of the $D_{\rm ver}$/$R_s$, where the HI fraction is defined as the ratio between HI gas mass ($M_{\rm HI}$) and stellar mass ($M_*$):
\begin{equation}
f_{\rm HI} = {\rm log} \frac{M_{\rm HI}}{M_{*}}.
\end{equation}
At a given $D_{\rm ver}$/$R_s$, lower-mass galaxies are systematically higher $f_{\rm HI}$ than higher-mass ones. This is consistent with the general relationship between $f_{\rm HI}$ and stellar mass found in the Virgo cluster, in which the average gas fraction increases with decreasing stellar mass \citep[see Figure 5 of][]{Hal12}.

The $f_{\rm HI}$ distribution of higher-mass galaxies is likely to show a trend with the $D_{\rm ver}$/$R_s$; $f_{\rm HI}$ decreases with increasing vertical distance from the filament spine. Lower-mass galaxies also appear to show a hint of a very smooth increase of $f_{\rm HI}$ with the $D_{\rm ver}$/$R_s$. However, the K-S tests of higher- and lower-mass galaxies yield probabilities of 0.2426 and 0.4351, respectively, for $f_{\rm HI}$ distributions between the inner ($0 \leq D_{\rm ver}/R_s \leq 1.75$) and outer ($1.75 < D_{\rm ver}/R_s \leq 3.5$) regions of the filament, indicating that $f_{\rm HI}$ gradients of both two subsamples are not statistically significant.

\section{Discussion} \label{sec:discussion}

In the filament structures around the Virgo cluster which are mostly composed of faint, low mass ({\lMsMs} $< 9$) galaxies, we observe negative stellar mass and $g-r$ color gradients (see Figure~\ref{ColorVDist}). These features are also shown in previous results based on the sample of massive galaxies in more distant filaments \citep{Mah18,Lub19,Sar19}; galaxies closer to the filament spine have higher stellar masses and redder colors than those showing large displacement from filaments. A generally suggested mechanism for this mass gradient is galaxy mergers in the past \citep{Mal17,Kra18}, in which more massive galaxies finish their assembly of stellar mass closer to the filament spine owing to a higher merger rate \citep{Dub14}. The formation of these massive merger remnants is accompanied by subsequent changes in their colors since the mass regulates properties of galaxies \citep{Pen10}. This is in line with the suggestion that mass is the main parameter driving galaxy properties in filaments \citep[e.g.,][]{Rob13,Alp15,Alp16}.

Interestingly, when we narrow down the mass ranges of galaxies by dividing into two subsamples (i.e., higher-mass with {\lMsMs} $> 8$ and lower-mass with {\lMsMs} $\leq 8$), mass gradients are not observed for both mass ranges (see bottom panel of Figure~\ref{MassVDist}). In this case, color gradients are also not anticipated in terms of the connection between stellar mass and color of galaxies. This is shown for higher-mass galaxies; stellar mass and color of higher-mass galaxies are almost statistically invariant with the vertical distance from the filament spine. However, in the case of lower-mass galaxies, the overall feature of the color distribution is decoupled from the stellar mass distribution; although stellar masses of galaxies are similar for all distances, the negative color gradient is clearly displayed (see top panel of Figure~\ref{MassVDist}). This indicates that an additional process might be responsible for the color gradient shown in lower-mass galaxies.

The local environment is generally considered as the second parameter in determining galaxy properties at fixed stellar mass \citep[e.g.,][]{Pen10,Geh12,Wet12}. In this regard, the interaction between galaxies is thought to be one of the typical environmental processes. The strong tidal forces in interacting galaxies can remove a large amount of gas from the outer part of galaxies, leading to a subsequent decrease in global star-formation activity \citep{Bar91,Mih94,Mih96,Mat07}. Meanwhile, galaxy interaction also causes funneling gas into the central region of galaxies, which could trigger strong central bursts of star-formation \citep{Bar96,Mih96,Kew06b}.

The $NUV-r$ color distribution of lower-mass galaxies shows a clear negative correlation with the distance from the filament spine (see Figure~\ref{SFGCD}), indicating global star-formation is depressed for lower-mass galaxies residing toward the filament spine. This is consistent with a general prediction that the efficiency of galaxy interaction could decrease moving further away from the filament spine. However, contrary to an expectation of gas flows toward the central region by galaxy interaction, lower-mass galaxies show no distinct variation of central star-formation, traced by EW(H$\alpha$), with the distance from the filament spine (see Figure~\ref{SFGHAEW}); i.e., lower-mass galaxies at closer to the spine do not show enhanced central star-formation. The restoring force of a lower-mass galaxy on possessing interstellar medium could be very small due to its shallow potential well. If the tidal force exerted by a companion galaxy is greater than the restoring force of the lower-mass galaxy, a large fraction of its gaseous material would be perturbed and then removed. Therefore, the EW(H$\alpha$) distribution of lower-mass galaxies could be explained by a lack of available fuel infalling from the outskirts of galaxies for their central star-formation.

In the case of higher-mass galaxies with deep potential wells, much of the original gas reservoir is likely to be retained against the tidal force by the interaction. Our result clearly indicates that global star-formation of higher-mass galaxies is not sensitive to the galaxies' distance from the filament spine (see Figure~\ref{SFGCD}). In this case, on the other hand, higher-mass galaxies residing closer to the filament spine would show an enhancement in central star-formation by funneling a large quantity of available gas into the central region of the galaxies (see Figure~\ref{SFGHAEW}).

The environmental mechanism related to the intergalactic medium such as ram-pressure stripping is important for regulating star-formation and the evolution of galaxies in galaxy clusters \citep[e.g.,][]{Gun72,Lar80}. \citet{Ben13} also show that hydrodynamical interaction between a galaxy and intergalactic gas in the filaments results in the removal of much of the gaseous component within a galaxy when the galaxy traverses the filaments. Therefore, ram-pressure stripping in the filament, the so-called ``cosmic-web stripping" \citep{Ben13}, is also expected to potentially influence on galaxy properties in the filament environment \citep{San08,Ara19}. Numerical simulations and X-ray observations might support this idea such that filaments host warm-hot intergalactic medium (WHIM) as the dominant baryon mass contribution in filaments \citep[e.g.,][]{Eck15,Aka17,Par17,Mar19,Tan19}.

The ram-pressure stripping can be probed by the HI cold gas content within galaxies. In this regard, \citet{Lub19} found a positive HI fraction gradient of galaxies with a mean stellar mass of $10^{10}$ $M_{\odot}$ in which galaxies further away from the filaments show a higher HI fraction, indicating that HI gas in galaxies closer to filaments can be efficiently removed by process related to the intrafilament medium. The deprivation of gas in galaxies would be more pronounced in low-mass galaxies due to their shallower potential wells. However, in the case of our lower-mass ({\lMsMs} $\leq 8$) subsample, the difference of HI fraction distributions between the inner and outer regions of the filament is not statistically significant (see Figure~\ref{HIFVDist}). This implies that the ram-pressure stripping does not play a significant role in reducing HI gas of lower-mass galaxies in filaments. It might be tempting to ascribe the lack of distinct HI fraction gradient of filament galaxies to the intrinsic property of intergalactic medium in the Virgo filaments characterized by a low density of WHIM.

In terms of understanding galaxy evolution in filaments, it can be said that galaxies can replenish HI gas content through cold mode gas accretion from the filaments (\citealt{Ker05} and references therein; \citealt{Kle17}). This is supported by numerical simulations which also show that filaments host diffuse HI gas at low redshift \citep[e.g.,][]{Pop09}. \citet{Kle17} found tentative evidence of cold gas accretion for massive galaxies with stellar masses of $>$$10^{11}$ $M_{\odot}$ from the intrafilament medium, whereas a hint of gas accretion was not observed for low-mass galaxies with $<$$10^{11}$ $M_{\odot}$. They proposed that cold mode gas accretion is only allowed for the most massive galaxies in filaments owing to larger gravitational potentials of galaxies enough to pull cold gas from the intrafilament medium. The difference in HI fraction of our higher-mass galaxies between the inner and outer regions of the filament is not found to be statistically significant (see Figure~\ref{HIFVDist}), indicating the absence of any gas accretion. Our result is in line with that of \citet{Kle17}, considering that the mass range of our higher-mass galaxies ({\lMsMs} $\sim$ 8--11) corresponds to that of their low-mass galaxies.


\section{Summary and Conclusions} \label{sec:summary}

In this study, we have explored properties of galaxies in six filaments around the Virgo cluster, which are the nearest large-scale structures from us ($\sim$14--41 Mpc), as a function of the vertical distance from the filament spine. Using the NSA and galaxy group catalogs, we selected galaxies that do not belong to galaxy groups in filaments and defined filament member galaxies that are located within 3.5 scale length from the filament spine. The main results are the following:

\begin{enumerate}[(i)]
\item The filaments are mainly composed of low-mass dwarf galaxies; the median value of the stellar mass distribution is {\lMsMs} $= 7.7$ and 87\% of the total sample have stellar mass with {\lMsMs} $< 9$. In $g-r$ and $NUV-r$ CMDs, the filament galaxies show a blueward offset from the red sequence of the Virgo cluster and are dominantly located on the blue cloud.

\item We observe that $g-r$ color and stellar mass correlate with the vertical distance from the filament spine; color becomes blue and stellar mass decreases with increasing vertical filament distance. We confirm that these trends are statistically significant in the K-S test between two subsamples located in the inner and outer regions of the filament. We propose that the negative color and stellar mass gradients of galaxies found in our filaments can be explained by mass assembly depending on efficiency of galaxy mergers at different vertical distances from the filament spine and subsequent change in their colors regulated by final mass.

\item We also examined the $g-r$ color and stellar mass distributions of galaxy subsamples in different mass ranges; higher-mass galaxies with {\lMsMs} $> 8$ and lower-mass galaxies with {\lMsMs} $\leq 8$. The $g-r$ color distribution of higher-mass galaxies is found to be different from that of lower-mass galaxies; higher-mass galaxies have a flat distribution against the vertical filament distance, whereas lower-mass galaxies show a clear negative color gradient. In the case of stellar mass distribution, a significant trend of stellar mass with the vertical filament distance is not shown for both subsamples in different mass ranges.

\item We inspected EW(H$\alpha$) and $NUV-r$ color distributions of higher- and lower-mass subsamples, which trace central and global star-formation activities of galaxies, respectively, using star-forming galaxies selected by the BPT emission line diagnositic diagram. The EW(H$\alpha$) distributions between two subsamples are different; we observe a negative EW(H$\alpha$) gradient for higher-mass galaxies, whereas lower-mass galaxies show no distinct variation of EW(H$\alpha$). On the other hand, $NUV-r$ color distributions exhibit counter-trend in different mass-ranges; the $NUV-r$ color distribution of higher-mass galaxies shows no statistically strong dependence on the vertical filament distance, but lower-mass galaxies show a distinct negative $NUV-r$ color gradient. All the observed EW(H$\alpha$) and $NUV-r$ color distributions of higher- and lower-mass subsamples can be qualitatively explained by different star-formation efficiency in galaxy interactions depending on masses of galaxies.

\item We observe no clear trend in the distributions of HI fraction for two subsamples detected from the ALFALFA blind survey data. Both higher- and lower-mass galaxies show no statistically significant variations in HI fractions with the vertical filament distance. This indicates that possible mechanisms related to intrafilament medium, such as ram-pressure stripping and gas accretion, could be ignorable for galaxies in the Virgo filaments. Although the ALFALFA data obtained from Arecibo single-dish telescope provides us a statistical view of the gas content of galaxies, further deep HI surveys using more sensitive very-large array telescopes will improve our understanding of the HI content of galaxies within the Virgo filaments.
\end{enumerate}

\acknowledgments
We are grateful to the anonymous referee for helpful comments and suggestions that improved the clarity and quality of this paper. Y.D.L, acting as the corresponding author, acknowledges support from Basic Science Research Program through the National Research Foundation of Korea (NRF) funded by the Ministry of Education (2020R1A6A3A01099777). S.C.R., acting as the corresponding author, acknowledges support from the Basic Science Research Program through the National Research Foundation of Korea (NRF) funded by the Ministry of Education (2018R1A2B2006445). Support for this work was also provided by the NRF to the Center for Galaxy Evolution Research (2017R1A5A1070354). S.K. acknowledges support from the Basic Science Research Program through the National Research Foundation of Korea (NRF) funded by the Research Staff Program (NRF-2019R1I1A1A01061237). We would like to thank Editage (www.editage.co.kr) for English language editing.


\bibliography{draft}
\end{document}